# The Stanford RNA Mapping Database for sharing and visualizing RNA structure mapping experiments

Pablo Cordero[1], Julius Lucks[2] and Rhiju Das[1,3,*]

[1]Program in Biomedical Informatics, Stanford University, Stanford CA 94305
[2] School of Chemical and Biomolecular Engineering, Cornell University, Ithaca NY 14853
[3]Departments of Biochemistry & Physics[4], Stanford University, Stanford CA 94305

* To whom correspondence should be addressed: rhiju@stanford.edu. Phone: (650) 723-5976. Fax: (650) 723-6783.

## Abstract

We have established an RNA Mapping Database (RMDB) to enable a new generation of structural, thermodynamic, and kinetic studies from quantitative single-nucleotide-resolution RNA structure mapping (freely available at http://rmdb.stanford.edu). Chemical and enzymatic mapping is a rapid, robust, and widespread approach to RNA characterization. Since its recent coupling with high-throughput sequencing techniques, accelerated software pipelines, and large-scale mutagenesis, the volume of mapping data has greatly increased, and there is a critical need for a database to enable sharing, visualization, and meta-analyses of these data. Through its on-line front-end, the RMDB allows users to explore single-nucleotide-resolution chemical accessibility data in heat-map, bar-graph, and colored secondary structure graphics; to leverage these data to generate secondary structure hypotheses; and to download the data in standardized and computer-friendly files, including the RDAT and community-consensus SNRNASM formats. At the time of writing, the database houses 38 entries, describing 2659 RNA sequences and comprising 355,084 data points, and is growing rapidly.

## Introduction

Understanding the secondary and tertiary structures of RNAs is critical for understanding their diverse biological functions, ranging from catalysis in ribosomal RNAs to gene regulation in metabolite-sensing riboswitches and protein-binding elements in RNA messages (1-7). RNA structure has therefore been intensely studied with a variety of biophysical and biochemical technologies (8-12). Amongst these tools, a facile, information-rich, and widely used class is chemical or enzymatic mapping. In these methods, a chemical probe or enzyme modifies or cleaves the RNA at susceptible nucleotides with a rate correlating with the exposure or flexibility of the site. Readout of modification or cleavage can be achieved through different means. For chemical modification, typically a reverse transcription step gives end-labeled DNA fragments whose length distributions can be resolved at nucleotide resolution by gel electrophoresis. These distributions can then be quantified and mapped back to the original sequence for RNAs as large as ribosomes (19, 28). In recent years, investigators have developed high-throughput technologies, such as 96-well capillary electrophoresis (13-14) or deep sequencing (15), to perform this step. Furthermore, several bioinformatic pipelines have been designed and implemented to rapidly quantify, map, and analyze the resulting data (16-18). RNA mapping experiments are now routinely used to improve automated secondary structure modeling (19), probe entire viral genomes (20), approximate secondary structures of whole yeast transcriptomes (30), and estimate an RNA's 'contact map' by coupling to exhaustive single-nucleotide mutagenesis (21-22).

The increase in throughput and the ease of quantification of these assays is leading to an accumulation of hundreds of thousands of quantitative data points per year that, taken together, can become a powerful resource for RNA structural characterization and further bioinformatic interrogation. This potential has been recognized in recent efforts to encourage researchers to create decentralized resources to share RNA mapping data using the proposed Single Nucleotide Resolution Nucleic Acid Structure Mapping (SNRNASM) Isa-TAB file format (23). In this model, researchers provide their data through a third party web application (e.g., Google Docs) and the community is then invited to annotate and validate it. However, SNRNASM does not provide researchers with data visualization tools or programming-friendly data formats to exploit the shared data for additional computational studies or meta-analysis. Furthermore, its distributed nature, the difficulty of enforcing standards in experimental protocols and file formats, and reliance on a third party hinders building web applications and robust meta-analysis tools that are needed for the next generation of RNA bioinformatics and structure determination. For these critical uses, a centralized resource with standardized and systematically curated data is needed. With this objective in mind, we have created the Stanford RNA Mapping Database (RMDB). The RMDB is a repository of RNA mapping experimental data and analysis that can be explored through a dynamic web interface, as well as downloaded in programming-friendly formats. In addition, the RMDB provides online tools and a software toolkit to visualize, manipulate, and analyze mapping data.

We describe herein the RMDB using a top-down approach, in the manner of a software functional specification: starting with the data conceptualization and visualization tools the database offers, followed by the database organization and content, and ending with implementation details.

## Data Visualization Front-end

For a specific RNA in a defined solution, each structure mapping experiment can be conceptualized as $M \times N$ matrices, where $M$ is the number of reactivity measurements made on the RNA, and $N$ is the number of nucleotides in the RNA. Entries in the Stanford RMDB house these data matrices, and are enriched with annotations and free text to describe associated content. In order to easily visualize and explore the data, we have developed a web application front-end, housed at http://rmdb.stanford.edu/. Through this interface, the user can browse the database content, perform full text search on the entries, and access documentation and RNA structure prediction tools without needing to register.

Each RMDB entry has its own details page, where its annotations, comments, associated publications, and authors are summarized. Furthermore, the entry's data matrix is displayed as a heat-map with more reactive residues given in darker shades of gray.  Rows of the matrix represent individual chemical mapping experiments and columns represent nucleotide positions. This visualization mimics the familiar gel bands of a classic electrophoresis experiment and can be used to easily spot differences between traces, reads, or peak areas. Bar plots are also provided for each experiment, and can be browsed using the arrows of the plot visualizer (see Figure 1). Secondary structure diagrams, generated by the RNAstructure software suite, guided by the chemical mapping data (19) are rendered using the VARNA visualizer when the user clicks each sample row. The accepted, 'native' structure of the RNA is also drawn if available. Except for the raw traces, each plot is rendered in real time as an SVG directly into the browser's canvas. This approach creates dynamic and interactive plots, and allows the user to inspect the raw data or preview a secondary structure for a particular sample (see Figure 1).

## Database content and structure

RMDB entries are organized and stored following their data matrix representation. Each entry in the RMDB contains values for normalized peak areas (such as those calculated by HiTRACE (17), CAFA (13), or ShapeFinder (16)) and optionally raw electrophoresis traces, or alternatively, maximum likelihood estimates (15) and number of reads for deep sequencing experiments. The RNA sequence being studied, its secondary structure, standardized annotations, and free text comments are also included. An RMDB entry may describe more than one experiment and can include error estimates if available.

Each entry is given an RMBD ID, chosen by the user during upload, which must be different from prior IDs. To facilitate human readability, the ID consists of three groups of alphanumeric characters separated by underscores: the first group has 6 characters that describe the probed RNA; the second of length 3 that describe the probe used; and the third is a four digit numeric identifier. For example,TRP4P6_SHP_0003 is an ID for the Tetrahymena group I intron P4P6 domain, probed using 2´-OH acylation (SHAPE) chemistry. RMDB entries also include version numbers, as they can be further updated with additional annotations and data.

Currently, the Stanford RMDB houses chemical footprints for a diverse collection of non-coding RNAs, probed under different solvent conditions. The database is able to take any kind of single-nucleotide-resolution mapping data. It currently includes experiments using base methylation by dimethyl sulfate (DMS); base adduct formation by 1-cyclohexyl-(2-morpholinoethyl)carbodiimide metho-p-toluene sulfonate (CMCT)' and selective 2' hydroxyl acylation with primer extension (SHAPE) with either *N*-methylisatoic anhydride (NMIA) or 1-methyl-7-nitroisatoic anhydride (1M7). The RNAs probed include riboswitches, ribozyme and ribosomal domains, and tRNAs featured in several published studies (15,21,22,24), as well as human-designed sequences accruing in the internet-scale RNA engineering project EteRNA (http://eterna.stanford.edu) (see Table 1). For almost all of the RMDB entries, the cDNA fragment separation and analysis steps were carried out by 96-well format capillary electrophoresis and the HiTRACE pipeline (17) respectively; two entries, describing data from (15), were read out through Illumina paired-end next-generation sequencing and processed using the SHAPE-Seq protocol (15,18). The RMDB currently houses 38 entries, describing 2659 RNA sequences and comprising 355,084 data points.

Users may also upload their own experimental data, in RDAT file format (see below). Following the precedent established by the Protein Data Bank, data submission requires registration and validation before public availability. In particular, once uploaded, an entry will be put on hold, to be curated, before it is made visible to the public. Afterwards, the new entry can be updated and managed through the RMDB administration system.

### Database schema and the RDAT file format

One of the goals of the Stanford RMDB is to facilitate the meta-analysis of RNA mapping experiments, allowing computational and structural biologists to infer RNA structural properties by combining the data from multiple assays and experiments. To this end, we have created the RNA Data (RDAT) text file format, an annotation-based specification that is both computer friendly and can be easily read by humans. The RMDB database schema closely follows the organization of RDAT files, making it easier to manage and expose each entry's data (see Figure 2).

RDAT files are composed of three main sections: the general section, the construct section, and the data section. The general section contains information about the RDAT specification version (RDAT_VERSION) used and serves as the root of the document. The construct section describes the specific RNA molecule that was probed in the experiment (NAME) and lists information about the construct, such as nucleotide sequence (SEQUENCE), putative secondary structure (STRUCTURE), solution conditions (ANNOTATION) and additional free text comments (COMMENTS). Additional optional fields can specify an integer offset that can be added to the RNA sequence index to yield a conventional numbering (OFFSET); the sequence positions to be specified in the file (SEQPOS); and the mutation positions of constructs in the data (MUTPOS), if there are any.

The mapping data for each construct is then encapsulated in data sections, with two required lines, ANNOTATION_DATA and REACTIVITY, for each lane/capillary of an electrophoresis experiments or for each sequence position in a deep sequencing experiment. We designed the ANNOTATION_DATA line format to give a concise, human-readable representation of experimental descriptions (e.g., the type of probes used, the ion concentrations, or if the data has been background subtracted). In particular, each space-separated entry in an ANNOTATION_DATA line gives attributes contained in name-value pairs (e.g., 'temperature:24C'). These annotations are hierarchical, in a way that a parent annotation (specified in the 'ANNOTATION' line in the previous general section) is inherited to all constructs (for example, modifier:CMCT given as an annotation for the general section would indicate that all constructs have been modified by CMCT). This hierarchy can also be overridden by more section specific annotations (in the example above, if there is a construct that had no modification, we place an annotation modifier:None to its construct section). A list of possible annotation types and their RDAT syntax is provided in table SI-1. This list completely describes the properties of experiments with different modifiers, purposes (structural characterization vs. thermodynamics), and experimental readouts; we add to this list as users deposit experiments with new modifiers.

The final required section of the RDAT file are the chemical modification data themselves, given in REACTIVITY lines. Sometimes, it may be desirable to summarize a set of experimental replicates in a single data section; errors derived from analysis of multiple sets of data can be reported under the REACTIVITY_ERROR data headers. After these key data are given, the file may contain (optional) additional raw data, including electropherogram traces (TRACE, such as in entry MDLOOP_SHP_000) and the positions in the trace that were called as peaks (XSEL).

Previously, an Isa-TAB compliant file format was proposed to store and share RNA mapping data (23). This format is perfect for manually inspecting the experimental results, as it is can be easily visualized in programs like Excel or Google Docs and only contains peak area or read count data information. For automatic analyses, however, the Isa-TAB format can be cumbersome and some experimental details, such as electropherogram traces, are lost. In contrast, the RDAT file format provides a more rigid and computer friendly structure, and keeps experimental details that can be useful when performing meta-analyses. Each format has a different goal, and therefore each RMDB entry can be downloaded in both formats. Data files deposited in either format are automatically converted to the other and sent through a validation tool.

### Additional tools and packages

In addition to enabling facile access to chemical mapping data, the RMDB provides several tools to handle RNA mapping data, including a secondary structure prediction server. This tool is available at http://rmdb.stanford.edu/structureserver/ and is mainly used to predict RNA secondary structure guided by experimental bonuses with the RNAstructure (v5.3) package (29). These bonuses can be uploaded in RNAstructure format, in an RDAT file, or contained in an RMDB entry. Bonuses can be either one-dimensional [such as the pseudoenergy bonuses described in (19)] or two dimensional [for mutate-and-map experiments, such as in (27)]. Other options, such as data normalization [as performed in (19)] and bootstrapping (26) are also included.

The server and database applications leverage several python modules and MATLAB scripts that serve as object oriented interfaces to popular secondary structure prediction algorithms and include parsers for the RDAT and Isa-TAB formats. We found that these software tools make it much easier to handle data contained in these formats and to perform secondary structure modeling and visualization. We therefore bundled them into a programmer's toolkit named the RDATkit, which can be downloaded from its SVN repository at https://simtk.org/svn/rdatkit. Both the structure server and the RDATkit are meant to facilitate the use of the data available in the RMDB in generating and testing structural hypothesis.

## Implementation details and availability

The paradigm of modern web applications, which mainly revolves around the Model-View-Controller concept, allows for rapid prototyping and implementation of database visualizers using standard development frameworks. In our case, we leveraged the power of the Django python web framework for server side logic, hosted by an Apache 2.2/MySQL 14.1 setup. The client-side visualization tools make use of JavaScript and SVG, supported by libraries such as jQuery, protovis, and D3. We find that this client-heavy approach allows for easy creation of dynamic and interactive visualizations that make full use of modern browser capabilities. To accelerate loading times, we also pre-generate some commonly used images in the visualizer and data browser (such as thumbnails and images for electropherogram traces), using a combination of the VARNA visualization tool (25), imagemagick, and python's matplotlib. The VARNA applet is also used in the client side to display RNA structures interactively.

All data can be browsed freely without registration. However, because of the use of native SVG to draw some of the graphics, a modern browser is required to fully explore the RMDB. The site has been tested on Firefox 4+, Internet Explorer 8 with the Google Chrome Frame or similar plugins installed, Chrome 6+, and Safari 5+. Code for the database and server implementation follows the GNU Public License and is available upon request.

## Discussion and Future Directions

We have presented the Stanford RMDB, a centralized and curated resource for RNA mapping experiments. The current data for 2395 RNA molecules will be of interest to structural biologists, bioinformaticians, and the RNA community in general. Currently, the RMDB holds data for mapping experiments that use chemical modifications to assess RNA structure. However, the database can be easily expanded to include experiments using enzymatic or hydroxyl cleavage. The size, and therefore utility, of the RMDB will greatly increase in the coming years. From the EteRNA project alone we obtain experimental results from 8-16 different sequences per week, and this rate is projected to rise. Further, the throughput of chemical and enzymatic mapping experiments has already taken a significant leap with multiplexed capillary electrophoresis, and next-generation sequencing (15) will allow probing of thousands of RNAs at once.

Rather than competing with other approaches of sharing mapping data, such as the decentralized SNRNASM repository, the RMDB completes the data's cycle as it provides a curated publishing endpoint as well as visualization and analysis tools.  Numerous questions can now be addressed, and new tools benchmarked: do modern vs. conventional chemical modifiers give better models? Are there automated and accurate ways to extract tertiary structure, in addition to secondary structure, information from these data?  Until recently researchers have not had the tools or the need to share high-throughput quantified RNA mapping data. It is therefore no surprise that thorough meta-analyses of these experiments are not yet available. It is our hope that the RMDB will make such projects possible and create resources to answer questions about RNA biology we have not yet formulated.

## Acknowledgements and Funding

We thank the authors of RNAstructure, ViennaRNA, VARNA, and the Stanford Visualization Team for making their source code freely available. We also thank R. Astorga for useful web design comments and critique as well as the members of the Das lab and C. VanLang for help with data curation and web page suggestions. This project was supported by the Burroughs-Wellcome Foundation (CASI to RD) and a Hewlett-Packard Stanford Graduate Fellowship (to PC). The server that houses the Stanford RMDB was graciously donated by K. Beauchamp.

## Tables

**Table 1: RNAs, solvent conditions, chemical modifiers, and types of experiments that are currently featured in the RNA Mapping Database.**

| RNA | Length | Chemical Modifiers | Solvent Conditions | Experiment types |
|---|---|---|---|---|
| *5S RNA, E. Coli* | 170 | NMIA | 10 mM MgCl$_2$, 50 mM Na-HEPES, pH 8.0 | Standard State, Mutate and Map |
| *adenine riboswitch, add* | 121 | NMIA | 10 mM MgCl$_2$, 50 mM Na-HEPES, pH 8.0, 0 and 5 mM adenine | Standard State, Mutate and Map |
| *cidGMP riboswitch, V. Cholerae* | 135 | NMIA | 10 mM MgCl$_2$, 50 mM Na-HEPES, pH 8.0, 0 and 10 uM ci-dGMP, | Standard State, Mutate and Map |
| *glycine riboswitch, F. nucleatum* | 198 | NMIA | 10 mM MgCl$_2$, 50 mM Na-HEPES, pH 8.0, 0 and 10 mM glycine | Standard State, Mutate and Map |
| *P4-P6 domain, Tetrahymena ribozyme* | 202 | NMIA | 10 mM MgCl$_2$, 50 mM Na-HEPES, pH 8.0, 0 and 30% methylpentanediol, | Standard State, Mutate and Map |
| *Ribonuclease P specificity domain, B. subtilis* | 198 | 1M7 | 10 mM MgCl$_2$, 100 mM NaCl, 100 mM Na-HEPES, pH 8.0 | Standard State |
| *SRP Domain IV* | 78 | DMS | 0–10 mM MgCl$_2$, 50 mM Na-HEPES, pH 8.0 | Titration |
| *tRNA phenylalanine (yeast)* | 116 | NMIA | 10 mM MgCl$_2$, 50 mM Na-HEPES, pH 8.0 | Standard State, Mutate and Map |
| *X20/H20* | 40 | DMS | 50 mM Na-HEPES, pH 8.0 | Mutate and Map |
| *MedLoop* | 80 | NMIA, DMS, CMCT | 50 mM Na-HEPES, pH 8.0 | Standard State, Mutate and Map |
| *One Bulge Cross* | 105 | NMIA | 10 mM MgCl$_2$, 50 mM Na- | Standard State |

|  |  |  | HEPES, pH 8.0 |  |

## Figures

Figure 1: Different ways to visualize entries in the RNA Mapping Database. (a) Classic bar graph of 2´-OH acylation rates across the nucleotides of the adenine-sensing domain of the *add* riboswitch from *V. vulnificus.* The data are from a 'standard state' study averaging 19 replicates across multiple experiments, and estimating the resulting errors (shown as error bars). (b) Exploring the contact map of the double-glycine-binding domain of a riboswitch from *F. nucleatum*. 2´-OH acylation data are shown for constructs with single mutations at each RNA position (the mutate-and-map approach). Inset: A user can select a particular modification lane and preview its experimentally constrained, predicted secondary structure.

**Figure 2  Data format and database schema.**(a) The RNA Data (RDAT) file format is designed to store RNA mapping data in a standardized manner. (b) An entity-relationship representation of the RMDB database schema, that is based on the RDAT file format.

a)

adenine riboswitch, add

**Mean values and error estimates for row 1 - SHAPE**

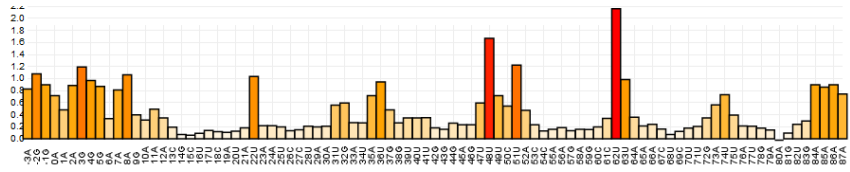

b)

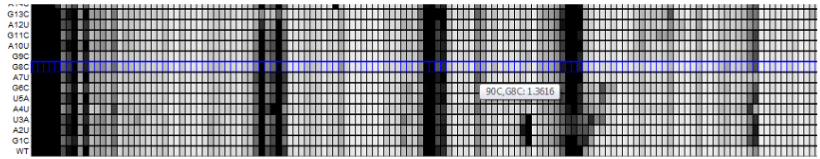

**Native structure**

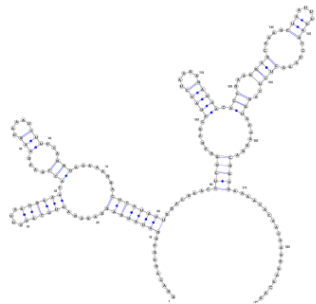
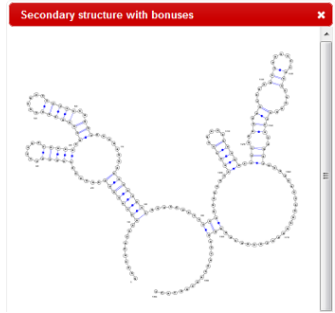

Secondary structure with bonuses

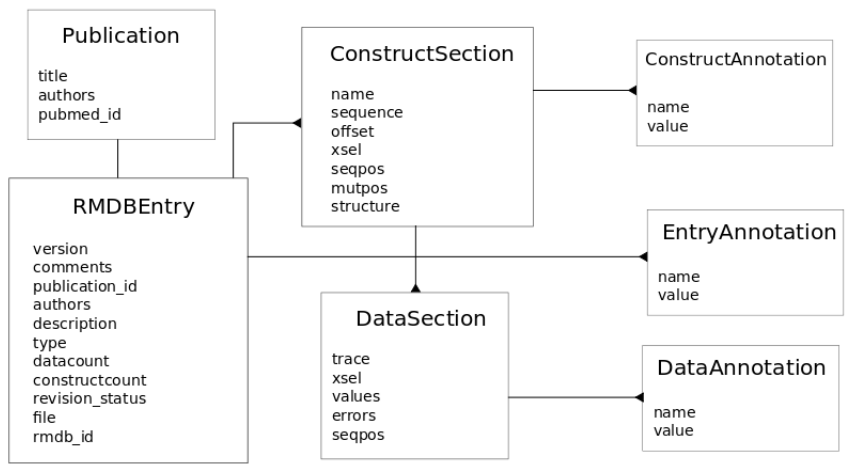

a)
```
RDAT_VERSION 0.22                                                    General section

NAME MedLoop                                                        Construct section
SEQUENCE  GGAACGACGAACCGAAAACCGAAGAAAUGAAGAAAGG...
STRUCTURE ..........(((((((((...............)))))))))...................
OFFSET -10
SEQPOS  50 49 48 47 46 45 44 43 42 41 40 39 38 37 36 35 34 33 32 31 ...
MUTPOS  WT  WT
ANNOTATION experimentType:StandardState chemical:Na-HEPES:50mM(pH8.0) temperature:24C
COMMENT   Example RDAT file with all fields filled: DMS and CMCT data for the MedLoop RNA.

                                                                       Data section
ANNOTATION_DATA    1      modifier:DMS
ANNOTATION_DATA    2      modifier:CMCT
AREA_PEAK          1      161.1038  70.2383  75.5198  88.3231  59.1168  44.1307  14.3177  15.7294 ...
AREA_PEAK          2      326.7871  50.8572  13.5739  24.2813   7.4603   2.3846  45.6323 263.8077
AREA_PEAK_ERROR    1       34.2208  16.0477  17.1040  19.6646  13.8234  10.8261   4.8635   5.1459 ...
AREA_PEAK_ERROR    2       67.3574  12.1714   4.7148   6.8563   3.4921   2.4769  11.1265  54.7615 ...
XSEL 93.00  126.00 158.00  187.00  214.00  238.00  260.00  279.00  299.00 ...
XSEL_REFINE        1       93.00   126.00  158.00  187.00  214.00  238.00  260.00  279.00 ...
XSEL_REFINE        2       90.00   119.00  154.46  186.00  212.00  234.68  260.88  281.26 ...
TRACE              1        0.0113   0.0050   0.0129   0.0278   0.0415   0.0674   0.0962 ...
TRACE              2        0.0245   0.0160   0.0098   0.0097   0.0247   0.0244   0.0124 ...
```

b)

## Supporting Information

Table SI-1: Some possible annotations, their respective SNRNASM terms, and RDAT used in the Stanford RMDB.

| Description of annotation | SNRNASM terms | RDAT syntax |
| --- | --- | --- |
| Modified sample using dimethyl sulfate | DMS:OBI:001015 | modifier:DMS |
| Modified sample using 1-cyclohexyl-(2-morpholinoethyl)carbodiimide metho-p-toluene sulfonate | CMCT:OBI:0001006 | modifier:CMCT |
| Modified sample using *N*-methylisatoic anhydride | NMIA:OBI:0001026 | modifier:NMIA, modifier:SHAPE |
| Modified sample using 1-methyl-7-nitroisatoic anhydride | 1M7:in process | modifier:1M7, modifier:SHAPE |
| Unmodified sample | - | modifier:None, modifier:Nomod |
| Chemical added to samples | - | chemical:*chemical name:concentration* |
| Secondary RNA sequence | - | chemical:*RNA* |

| added to samples | | *sequence:concentration* |
|---|---|---|
| RNA sequence changed in sample | - | sequence:*RNA sequence* |
| Specify temperature used | - | temperature:*temperature in celsiusC* |
| Specify processing method or algorithm used (such as background substraction or overmodification correction) | - | processing:*method name* |
| Indicate that the sequence was mutated, from N to M at nucleotide i (1-based sequence indexing) | - | mutation:NiM |
| Indicate that the sequence was mutated from NNN to MMM at nucleotide range i to j (1-based indexing) | - | mutation:NNN(i-j)MMM |
| Specify experiment type (currently supported: standard state, mutate and map, and titration) | - | experimentType:StandardState, experimentType:MutateAndMap, or experimentType:Titration, |